\documentclass[preprint]{aastex}

\usepackage{bm}
\usepackage{amsmath,amssymb,graphicx,natbib}
\usepackage[dvips]{color}

\begin{document}

\title{Rapid dynamical mass segregation and properties of fractal star clusters}

\author{Jincheng Yu\altaffilmark{1,2},
Richard de Grijs\altaffilmark{2,3}, Li Chen\altaffilmark{1}}

\altaffiltext{1}{Shanghai Astronomical Observatory, Chinese Academy of
Sciences, 80 Nandan Road, Shanghai 200030, P. R. China}
\altaffiltext{2}{Kavli Institute for Astronomy and Astrophysics, Peking University,
Yi He Yuan Lu 5, Hai Dian District, Beijing 100871, P. R. China}
\altaffiltext{3}{Department of Astronomy, Peking University, Yi He Yuan Lu 5, Hai
Dian District, Beijing 100871, P. R. China}

\begin{abstract}
We investigate the evolution of young star clusters using $N$-body
simulations. We confirm that subvirial and fractal-structured clusters
will dynamically mass segregate on a short timescale (within 0.5 Myr).
We adopt a modified minimum-spanning-tree (MST) method to measure the
degree of mass segregation, demonstrating that the stars escaping from
a cluster's potential are important for the temporal dependence of
mass segregation in the cluster. The form of the initial velocity
distribution will also affect the degree of mass segregation. If it
depends on radius, the outer parts of the cluster would expand without
undergoing collapse. In velocity space, we find `inverse mass
segregation,' which indicates that massive stars have higher
velocity dispersions than their lower-mass counterparts.
\end{abstract}


\section{Introduction}

Most stars with masses $m \ge 0.5 M_\odot$ are thought to form in star
clusters \citep{Lada}. It is, therefore, very important to understand
star formation in the context of star cluster formation. Observations
of very young and forming star clusters imply dynamically cool and
clumpy initial conditions \citep{Williams1999,Carpenter2008}.

Young clusters ($\lesssim6$ Myr) also often exhibit `mass
segregation,' where the massive stars are more centrally concentrated
compared to their lower-mass companions. The origin of this mass
segregation has been suggested as either `primordial,' that is, it is
a result of the star-formation process in which stars
\citep[particularly the massive stars;][]{Moeckel2010} form
mass-segregated from their parent molecular cloud
\citep{Chen2007,Weidner2010}, or dynamical, i.e., resulting from fast
dynamical evolution, with increased importance in the denser cores.

\cite{Allison2009} investigated the evolution of initially subvirial
and highly substructured open-cluster-like objects. They found that
this substructure will disappear on very short timescales (a few Myr),
with clusters undergoing mass segregation. Subsequently, these authors
\citep{Allison2010} also found that the more clumpy and cooler the
cluster is, the higher the degree of mass segregation will
be. However, not only the spatial distribution but also the stellar
velocity distribution affects the degree of mass segregation:
initially subvirial clusters will exhibit gradients in their radial
velocity structures. \cite{Proszkow2009apjs} found the initial
subvirial state of the star-forming clumps also plays an important
role in the sense that it facilitates high interaction rates, which
leads to rapid mass segregation.

In this paper, we investigate the influence of different initial
velocity distributions on the development of mass segregation. In \S2,
we introduce our initial conditions. In \S3, we analyze our
simulations, and in \S4, we discuss the results and draw our
conclusions.

\section{Method and initial conditions}

\subsection{Initial cluster model}

Both observations and theory suggest that young star clusters form
with subvirial kinematics and highly substructured in space and
velocity
\citep{Larson1995,Williams1999,Elmegreen2000,Testi2000,Cartwright2004,Carpenter2008,Schmeja2008,Proszkow2009apjs}.
We simulated the evolution of cool, clumpy star clusters containing
1000 single stars and no gas with initial virial radii $r_{\rm v}=0.5$
pc. The stellar initial mass function follows a three-part power law
after \cite{Kroupa2002}, with minimum and maximum masses of 0.08 and
$50M_\odot$, respectively. We follow the method of \cite{Goodwin2004}
to generate `fractal' clusters. First, we define a cube with sides
$N_{{\rm div}}$. We place the first-generation `parent' in the center
of the cube. Next, we split the cube into $N_{{\rm div}}^{3}$ parts,
each containing a subnode (`child') in its center. We adopt $N_{{\rm
div}}=2$ throughout, so that there will be eight subcubes and eight
children for each parent cube.

A child has a probability of $N_{\rm div}^{(D-3)}$ to become a parent
of the next generation, where $D$ is the fractal dimension. The lower
$D$ is, the fewer children become next-generation parents, and the
clumpier the cluster will be ($D=3.0$ corresponds to a uniform
distribution). After removing the first-generation parents and the
children that do not mature, we add Poissonian noise to the positions
of the parents to prevent the cluster from developing an obvious and
artificial gridded structure. Each mature child then divides into
$N_{\rm div}^{3}$ children in the centers of $N_{\rm div}^{3}$
subsubcubes. This process is repeated until there are many more
children than required. Eventually, cluster stars are selected
randomly from the remaining children. In this paper, we will not
discuss effects associated with varying the fractal dimension, so we
adopt a fixed fractal dimension $D=2.0$ to satisfy observational
constraints \citep{Montuori1997,Sanchez2009}.

\cite{Goodwin2004} also suggest that the velocity dispersion must be
coherent, which means that the children inherit their parents'
velocities. A random velocity component is also added to each child.
We argue that for star formation in molecular clouds, the velocities
of the stars depend on the potential they are embedded in, in the
sense that the stars in the inner parts should have higher velocities.
\cite{Kupper2010} show that the radial velocity-dispersion profiles
they derived based on a set of $N$-body simulations decline as a
function of radius for all bound stars as well as for the cluster
member stars within the Jacobi radius (where the internal
gravitational acceleration equals the tidal acceleration from the host
galaxy).

On the other hand, \cite{Rochau2010} use relative proper motions to
show that in the young ($\sim 1$ Myr) Galactic starburst cluster NGC
3603, stars with masses between 1.7 and $9M_\odot$ all exhibit the
same velocity dispersion, which implies that the cluster stars have
not yet reached equipartition of kinetic energy. Inspired by these
constraints, we will therefore consider two different forms of the
radial velocity-dispersion profile in our $N$-body calculations, i.e.,
a constant and a radially declining functional form. Although
molecular clouds do not possess a single dense core, there still
exists a dynamical center surrounded by many molecular-cloud
clumps. Thus, we define our radii with respect to this dynamical
center. The velocity distribution is then written as
\begin{equation}
v=v_{\rm parent}+\sigma,
\end{equation}
\begin{equation}
\sigma={\rm constant} \mbox{ or }
\sigma=\frac{\sigma}{1+(r/r_{\rm v})^2},
\end{equation}
where $v_{\rm parent}$ is the velocity of the parent node and $\sigma$
the velocity dispersion. Since the radial form of the nonconstant
initial velocity distribution is not well understood, we adopt a
plausible functionality that leads to a large velocity dispersion,
which in turn facilitates our exploration of its effects. We will
discuss both kinds of velocity dispersion in \S 3.

We only follow the evolution of our simulated clusters for a very
short time ($\le 6$ Myr), so that no stellar evolution is
included, given that its effect would be negligible on these
timescales. We assume, following \cite{Allison2009,Allison2010} and
for computational reasons, that our simulations commence at the time
that all stars have formed. We do not include the effects of the
surrounding gas to reduce adding to the complexity of the physics
involved; we aim at doing so in a subsequent paper.

Finally, for clarity we summarize the other parameters used for our
fractal model:
\begin{enumerate}
\item Total number of stars, $N=1000$; all stars are treated as single
stars;
\item Fractal dimension, $D=2.0$;
\item Initial virial ratio, $q=0.30$ (here defined as the ratio of
kinetic to potential energy);
\item Virial radius, $r_{\rm v}=0.5$ pc.
\end{enumerate}

We generate 100 samples for every velocity-distribution model. These
samples only vary with the initial random seed. This way, we can
address the resulting cluster properties statistically. We used {\tt
kira} in {\sc starlab} \citep{Zwart2001} as our $N$-body integrator.

\subsection{Measuring mass segregation}

We use the minimum-spanning-tree (MST) method \citep{Allison2009b} to
quantify the degree of mass segregation. An MST connects all sample
points using the shortest path length, without forming any closed
loops. There can be multiple shortest paths, but the length is
unique. We measure the degree of mass segregation of the $N$ most
massive stars by comparing their MST length with the average MST
length of $N$ randomly selected stars in the cluster (where both
values of $N$ are the same).

\cite{Allison2009b} define the amount of mass segregation by the ratio
of the random to the massive-star MST lengths,
\begin{equation}
\Lambda=\frac{\langle l_{\rm norm} \rangle}{l_{\rm massive}}\pm
\frac{\sigma_{\rm norm}}{l_{\rm massive}},
\end{equation}
where $\Lambda$ is the level of mass segregation (the
`mass-segregation ratio'), $l_{\rm massive}$ is the MST length of the
massive stars, and ${\langle l_{\rm norm} \rangle}$ and $\sigma_{\rm
norm}$ are the average length of and statistical error associated with
the MSTs of the random set, respectively. If $l_{\rm massive}$ is
significantly shorter than $\langle l_{\rm norm} \rangle$ (i.e., by
more than $\sigma_{\rm norm}$), the massive stars are more
concentrated and, therefore, the cluster is mass segregated.

However, it is not mathematically convenient to determine the average
$\Lambda$ from a set of simulations, because $l_{\rm massive}$ and
$\langle l_{\rm norm} \rangle$ from different samples may be
characterized by different weights when averaging the value of
$\frac{\langle l_{\rm norm} \rangle}{l_{\rm massive}}$ (i.e.,
$\Lambda$). That is, a sample with very large $\Lambda$ may dominate
the results. One reasonable approach is to average $\log(\Lambda)$
values instead of $\Lambda$.

Therefore, we use a new variable ($\Lambda'$) to quantify the degree
of mass segregation,
\begin{equation}
\Lambda'=\log \Bigl(\frac{\langle l_{\rm norm} \rangle}{l_{\rm
    massive}} \Bigr)\pm \log \Bigl(\frac{\sigma_{\rm norm}}{l_{\rm
    massive}} \Bigr).
\end{equation}
We will show the benefits of using the new variable in \S3.

\section{Analysis and results}

A cluster loses its member stars through energy exchange and tidal
effects \citep{Conversea}. In general, we are concerned about a
gravitationally consistent sample, that is, we cannot compare a
simulated cluster with several stars at great clustercentric distances
to a real cluster. Therefore, physically speaking, it is necessary to
remove the stars that do not belong to the cluster. We will show that
this effect is not negligible.

The evolution of the velocity distribution is also very important for
understanding the dynamical evolution of the cluster. Therefore, we
will discuss the influence of the choice of the initial velocity
distribution. In addition, we will also explore the behavior of
massive stars in velocity space.

\subsection{Effect of escaping stars}

Removing the escaping stars from the cluster is physically necessary.
On the other hand, there could also be some potential escapers that
may significantly affect the cluster's subsequent evolution
\citep{Kupper2010}. However, the amount of mass segregation is a
property of a cluster at a fixed time. Therefore, it is necessary to
calculate the amount of mass segregation after removing the escapers
but all member stars (no matter whether they are potential escapers)
should be included at any given time step. From this point of view, we
only need to truncate the cluster at a specific radius. In the
following, we will discuss the influence of our treatment of escaping
stars.

First, we compare the temporal dependence of the mass-segregation
ratio without removing the escaping stars for both $\Lambda$ and
$\Lambda'$.
%
\begin{figure}
  \begin{center}
    \includegraphics[width=6cm,angle=270]{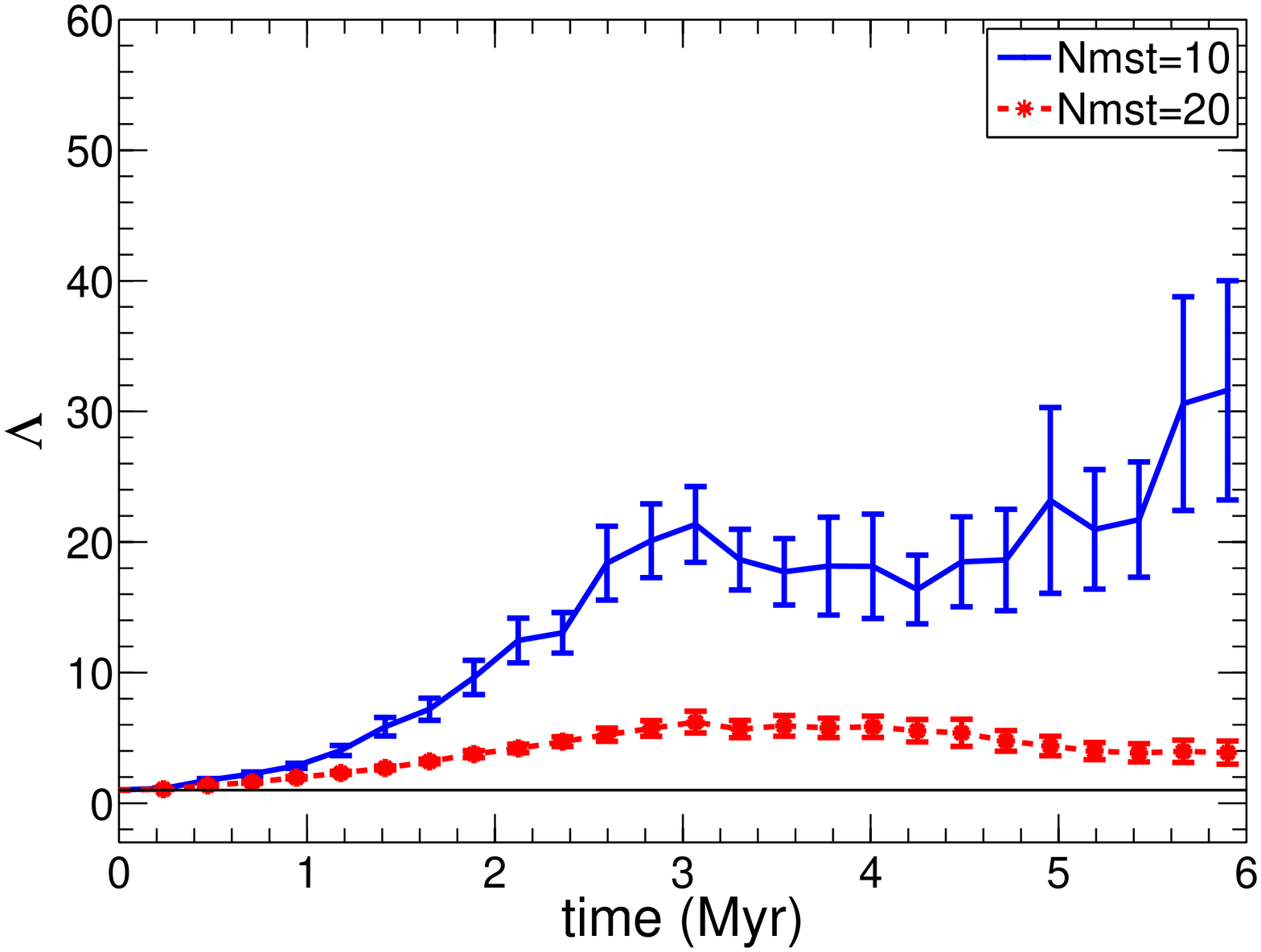}
    \includegraphics[width=6cm,angle=270]{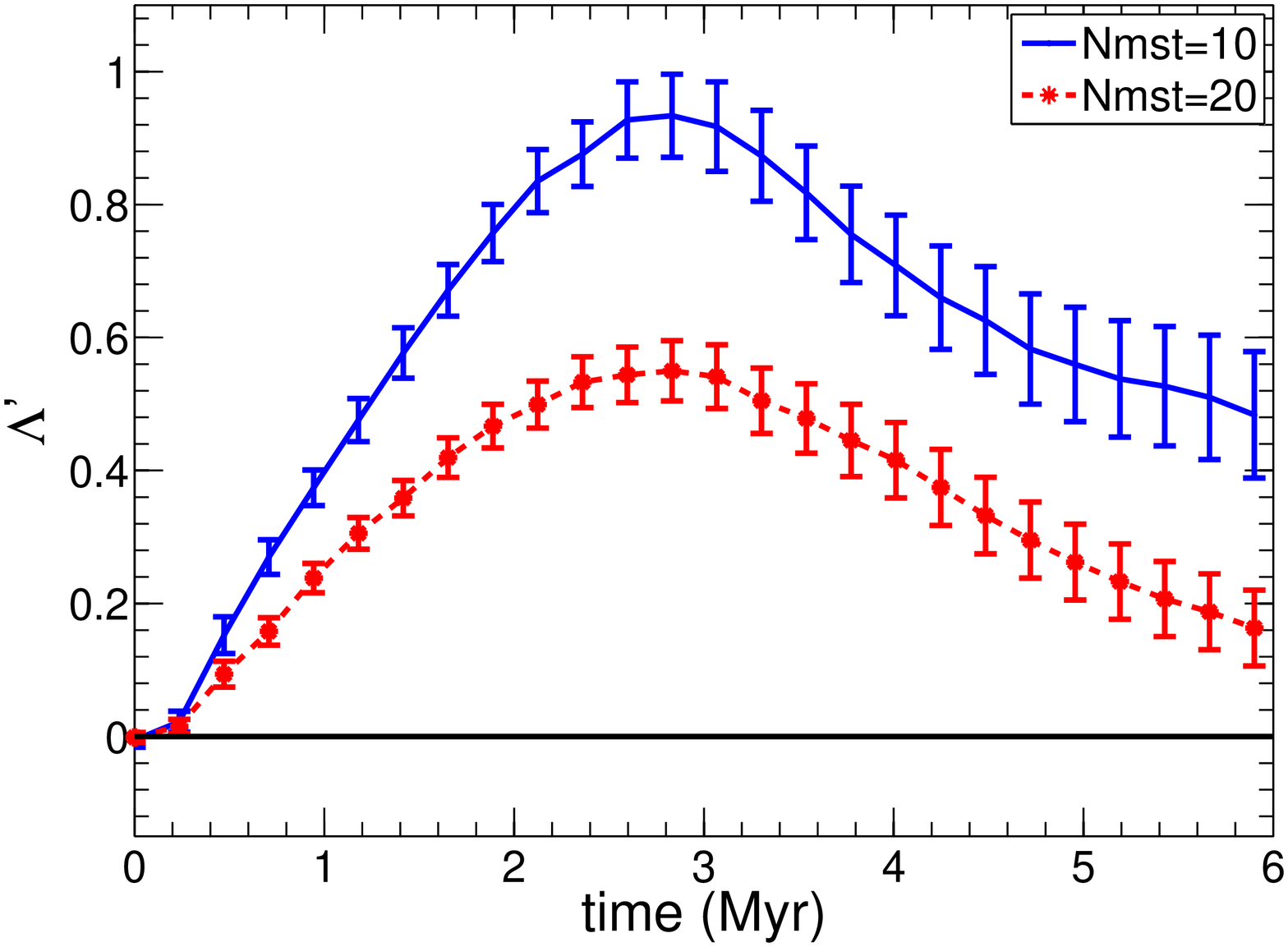}
  \end{center}
  \caption{Evolution of the mass-segregation ratio for the 10 and 20
    most massive stars (Nmst) using $\Lambda$ (left) and $\Lambda'$
    (right). The line $\Lambda=1$ (or $\Lambda'=0$) represents the
    no-mass-segregation situation.
  \label{fig:Lambda-t-before-rm}}
\end{figure}
%
We use average $\Lambda$ and $\Lambda'$ values for a given set of
samples. Note that the average of $\log(\Lambda)$ is not equal to the
logarithm of the average of $\Lambda$. The left-hand panel of
Fig. \ref{fig:Lambda-t-before-rm} shows an increasing degree of mass
segregation of the cluster with time from a very early stage. We also
note that the dispersion of $\Lambda$ increases with time, which
suggests that some $\Lambda$s may be extremely large, while others can
be less than unity (i.e., the massive stars are less concentrated than
the randomly selected sample stars). Therefore, if $\Lambda$ has a
large dispersion, this may lead to an incorrect result, as we
confirmed using $\Lambda'$. There is a turnover point at approximately
3 Myr, shown in the right-hand panel of
Fig. \ref{fig:Lambda-t-before-rm}. That is, the cluster becomes more
and more mass segregated until an age of 3 Myr and then the degree
decreases. This turnover point may be related to our treatment of the
escapers from the cluster. After truncating the cluster at a radius of
2.5 pc ($5r_{\rm v}$), we analyze the massive stars and the total
number of member stars in the cluster.
%
\begin{figure}
  \begin{center}
    \includegraphics[width=6cm,angle=270]{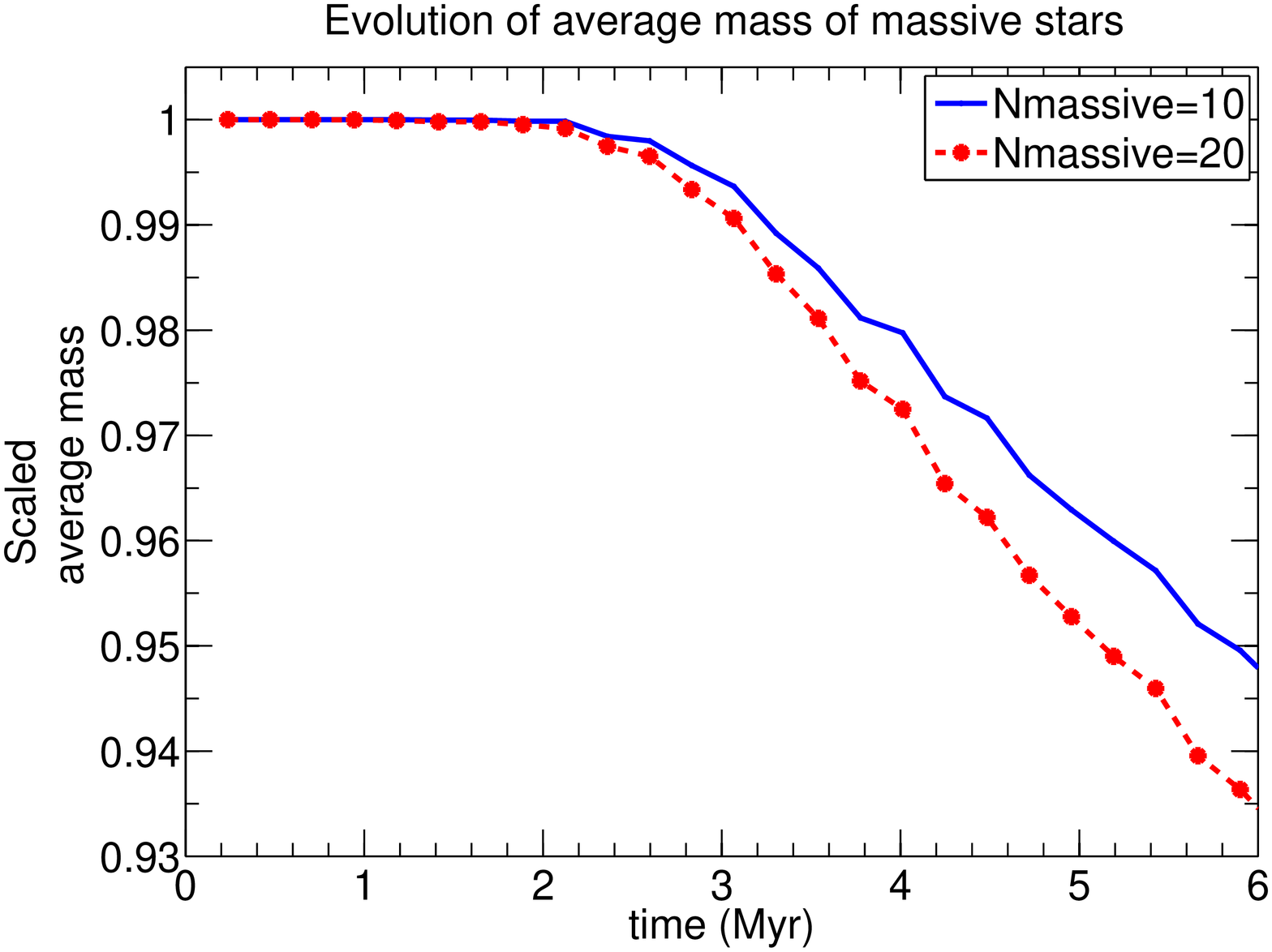}
    \includegraphics[width=6cm,angle=270]{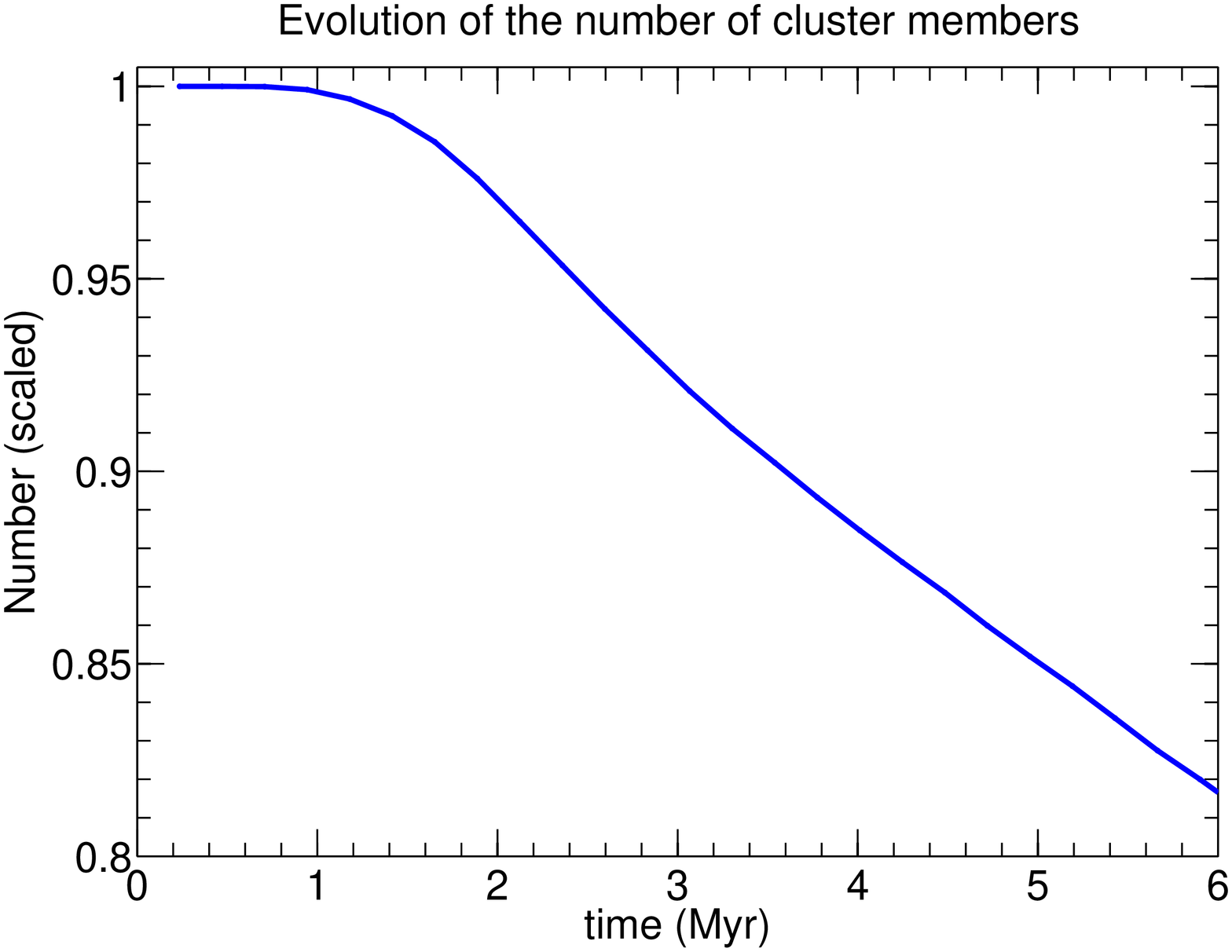}
  \end{center}
  \caption{Evolution of the average mass of the 10 and 20 most massive
    stars (left) and the total number of stars in the cluster after
    truncating the cluster at a radius of $5 r_{\rm v}$ (2.5 pc;
    right).
  \label{fig:mass-loss}}
\end{figure}
%
Fig. \ref{fig:mass-loss} shows the average mass of the 10 and 20 most
massive stars and the total number left in the cluster as function of
time. The average mass of the massive stars begins to decrease almost
at the same time as the turnover point develops, while the cluster
starts to lose members a little ($\le 1$ Myr) earlier. When some of
the massive stars leave the cluster, the MST length that includes
these escapers becomes very large, leading to a relatively small value
of $\Lambda'$. This is the reason for the turning point in the
$\Lambda'$--$t$ curve. Therefore, we will radially truncate our
clusters in the remainder of this paper.

\subsection{Choice of velocity distribution}

\begin{figure}
  \begin{center}
    \includegraphics[width=6cm,angle=270]{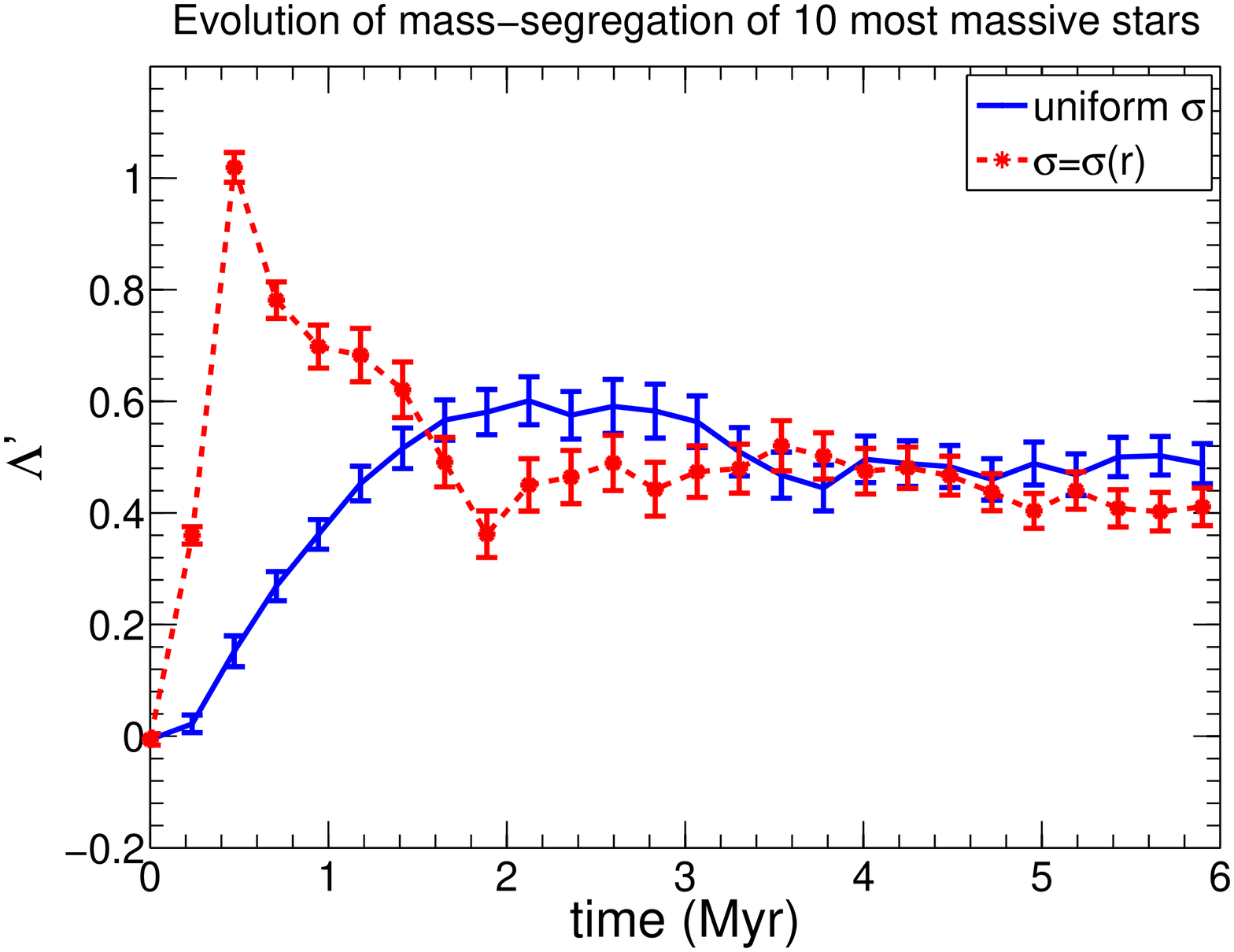}
    \includegraphics[width=6cm,angle=270]{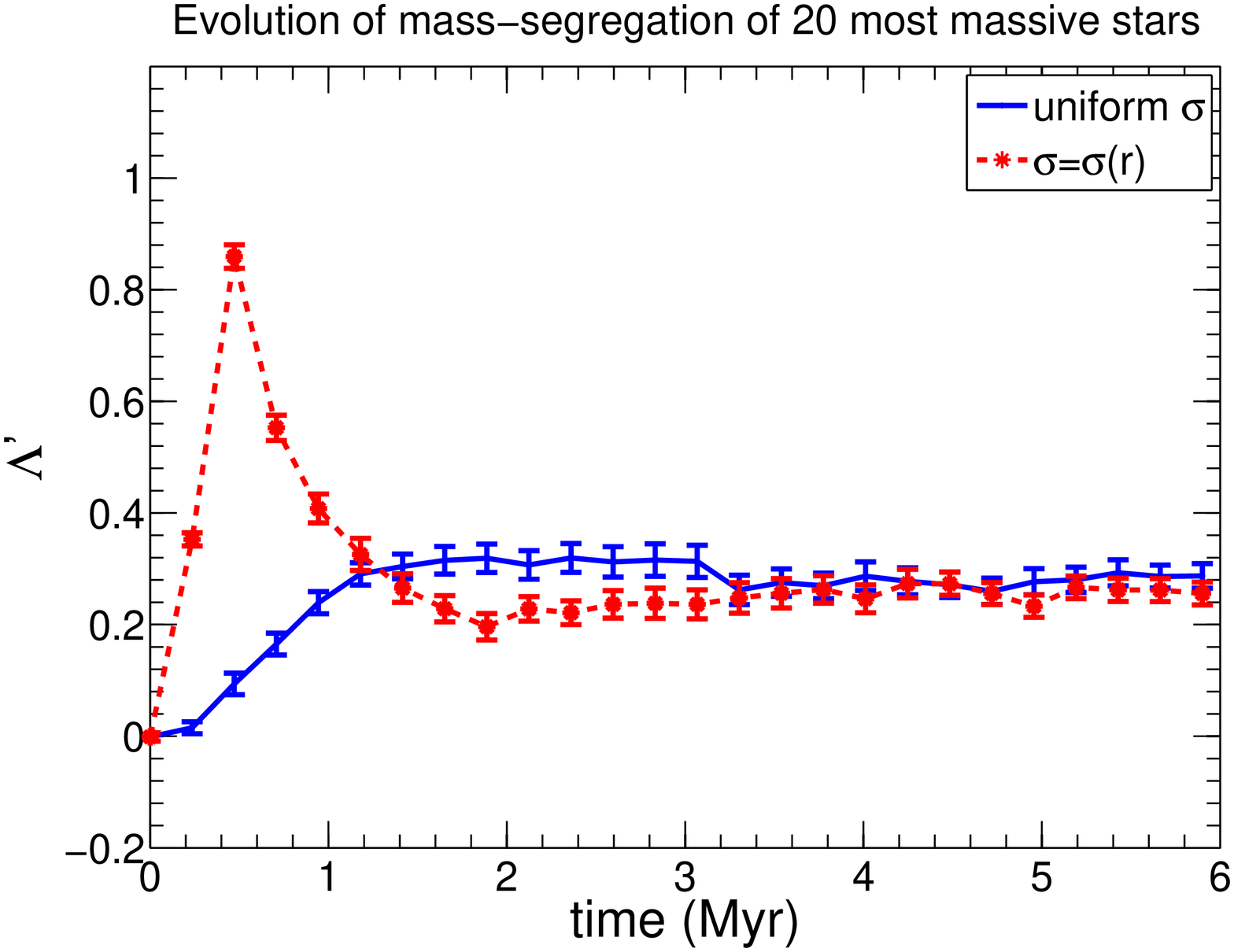}
  \end{center}
  \caption{Evolution of the (left) 10 and (right) 20 most massive
    stars in the cluster for uniform and radius-dependent velocity
    dispersions.
  \label{fig:Lambda-t}}
\end{figure}

Because clumpy clusters evolve rapidly, small changes may
significantly affect the results on short timescales. We find that the
$\Lambda'$ profile (as a function of time) changes if we vary the
velocity distribution. In Fig. \ref{fig:Lambda-t}, we compare the
results of the two different initial velocity distributions from
Eq. (2). One is uniform throughout, while the other is radius
dependent. For the radius-dependent velocity distribution, the degree
of mass segregation reaches a peak at 0.5--1 Myr. Shortly thereafter,
it drops to a relatively low level (similar to that of a uniform
distribution). We will now attempt to offer an explanation for this
behavior.
%
%

\begin{figure}
  \begin{center}
    \includegraphics[width=6cm,angle=270]{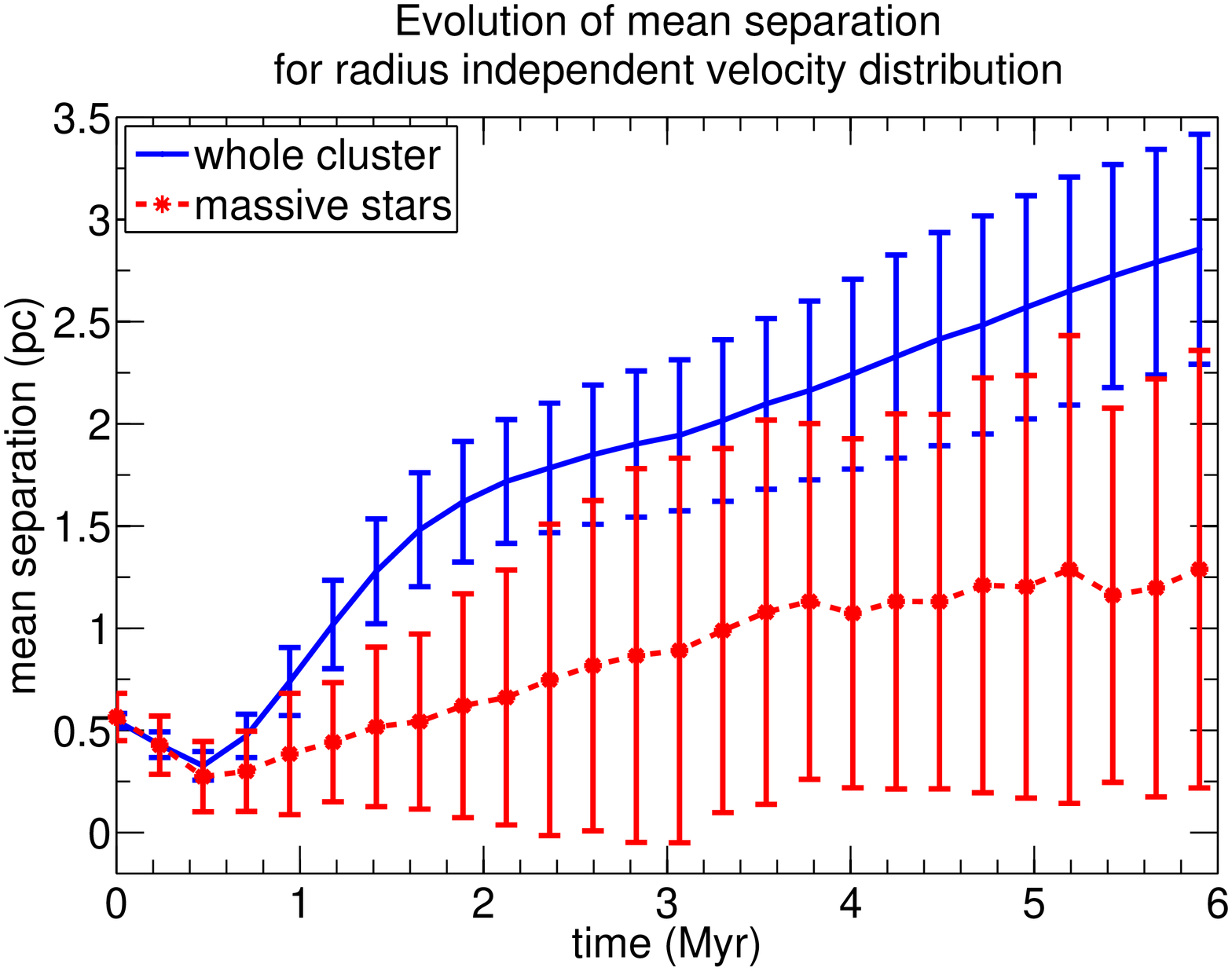}
    \includegraphics[width=6cm,angle=270]{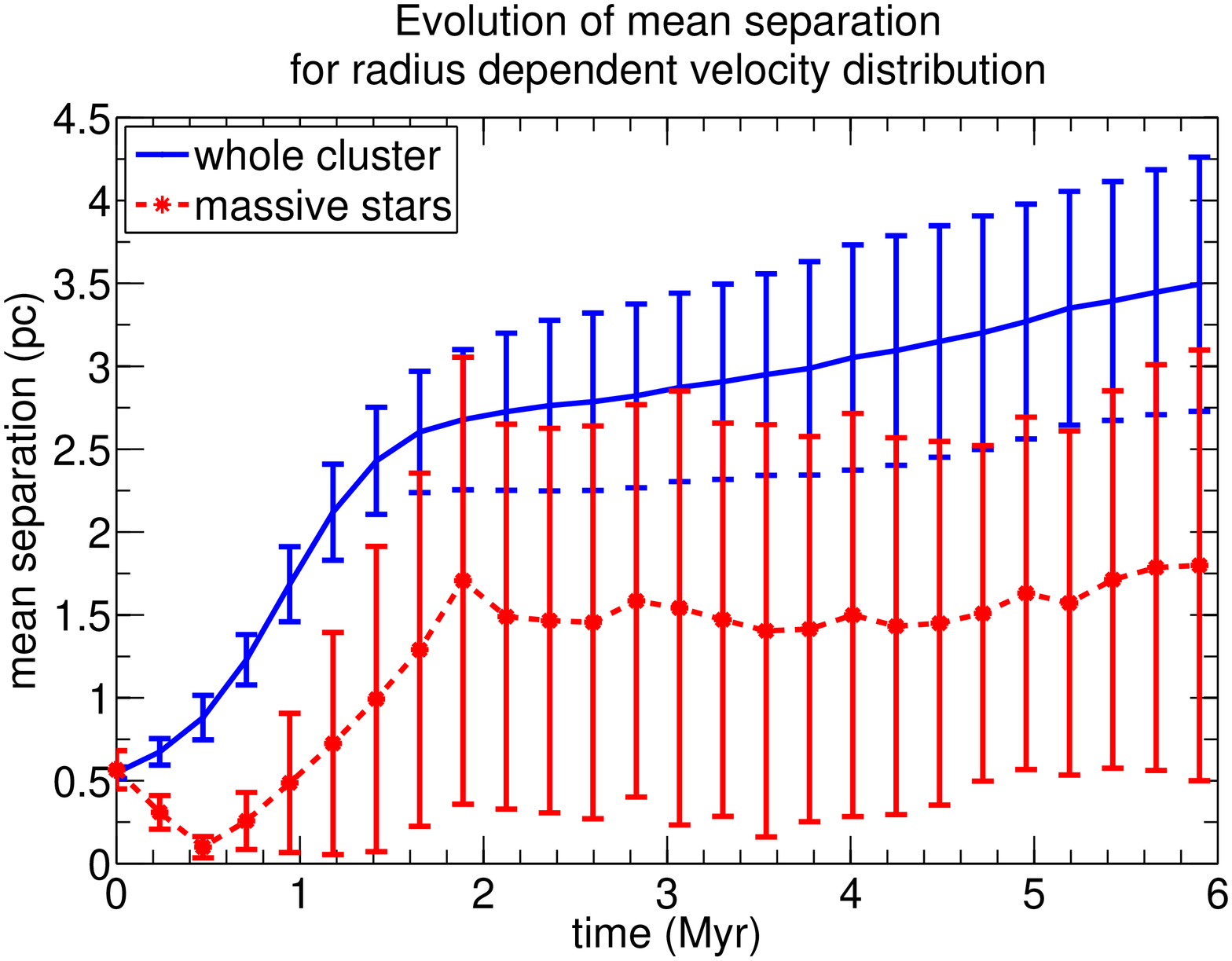}
  \end{center}
  \caption{Evolution of the mean separation for (left) uniform and
    (right) radius-dependent velocity dispersions. The solid line is
    the mean separation in the entire cluster, while the dashed line
    represents the mean separation of the 10 most massive stars in the
    cluster.
  \label{fig:MST-t}}
\end{figure}

\begin{figure}
  \begin{center}
    \includegraphics[width=6cm,angle=270]{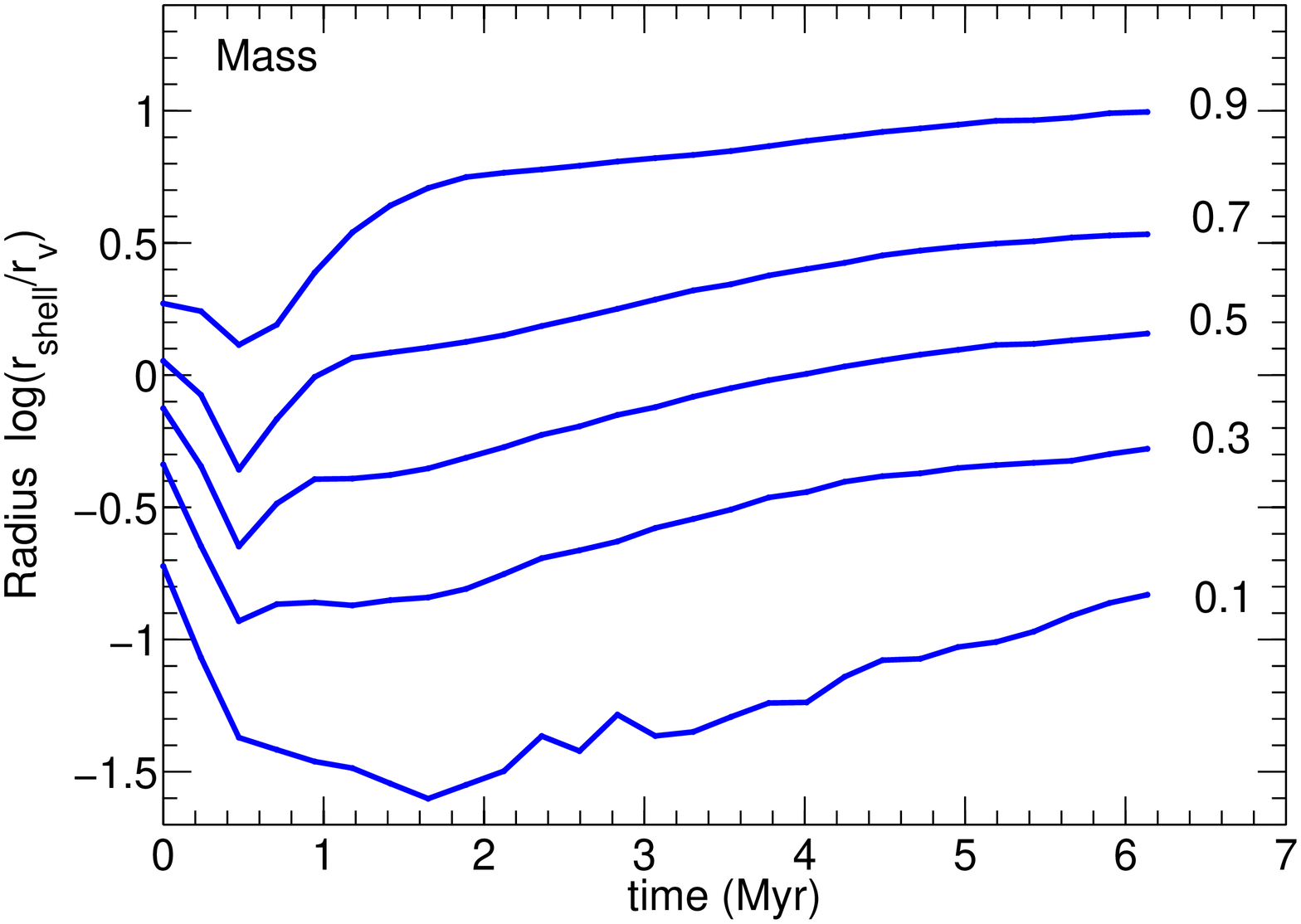}
    \includegraphics[width=6cm,angle=270]{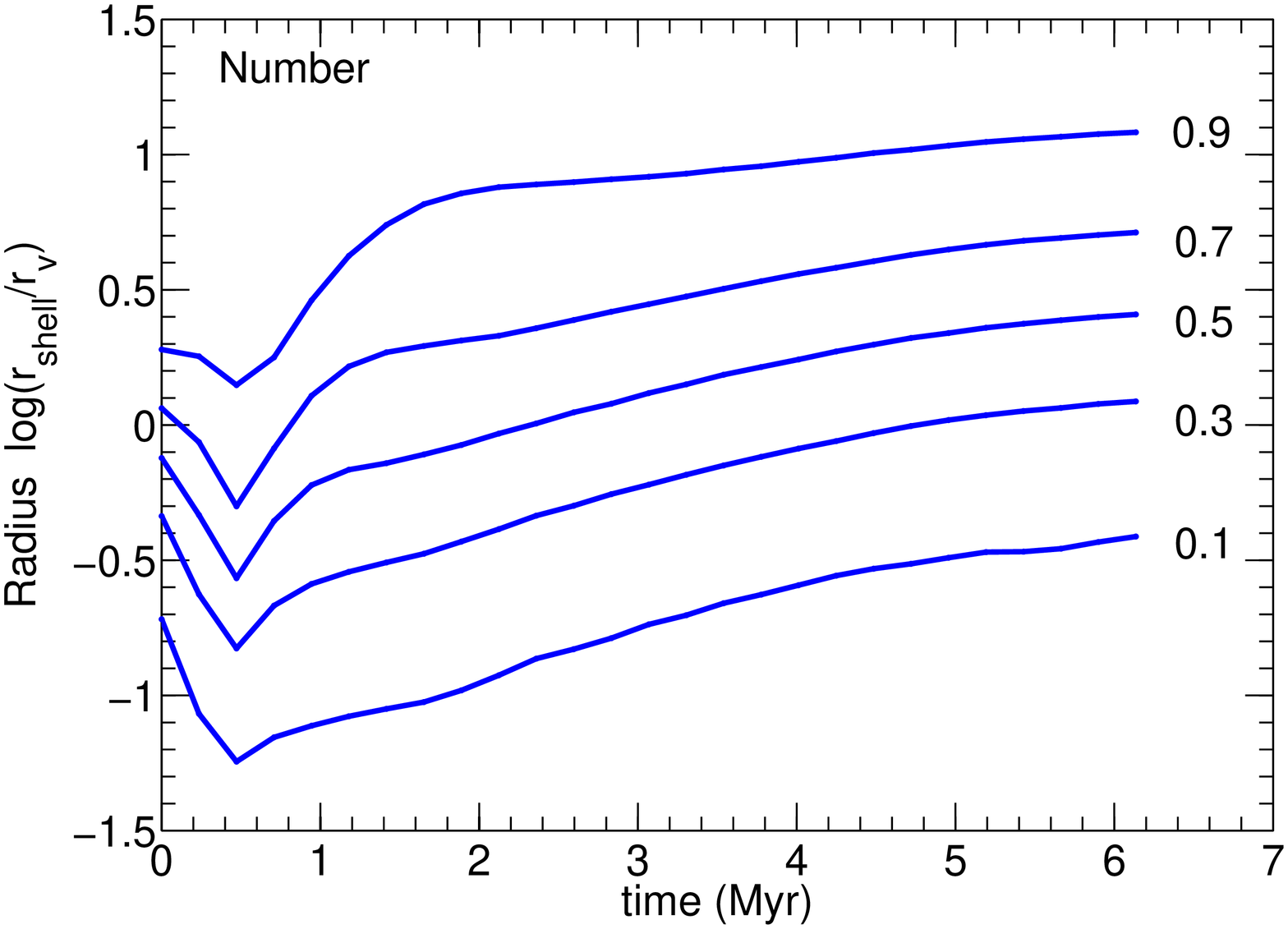}
    \includegraphics[width=6cm,angle=270]{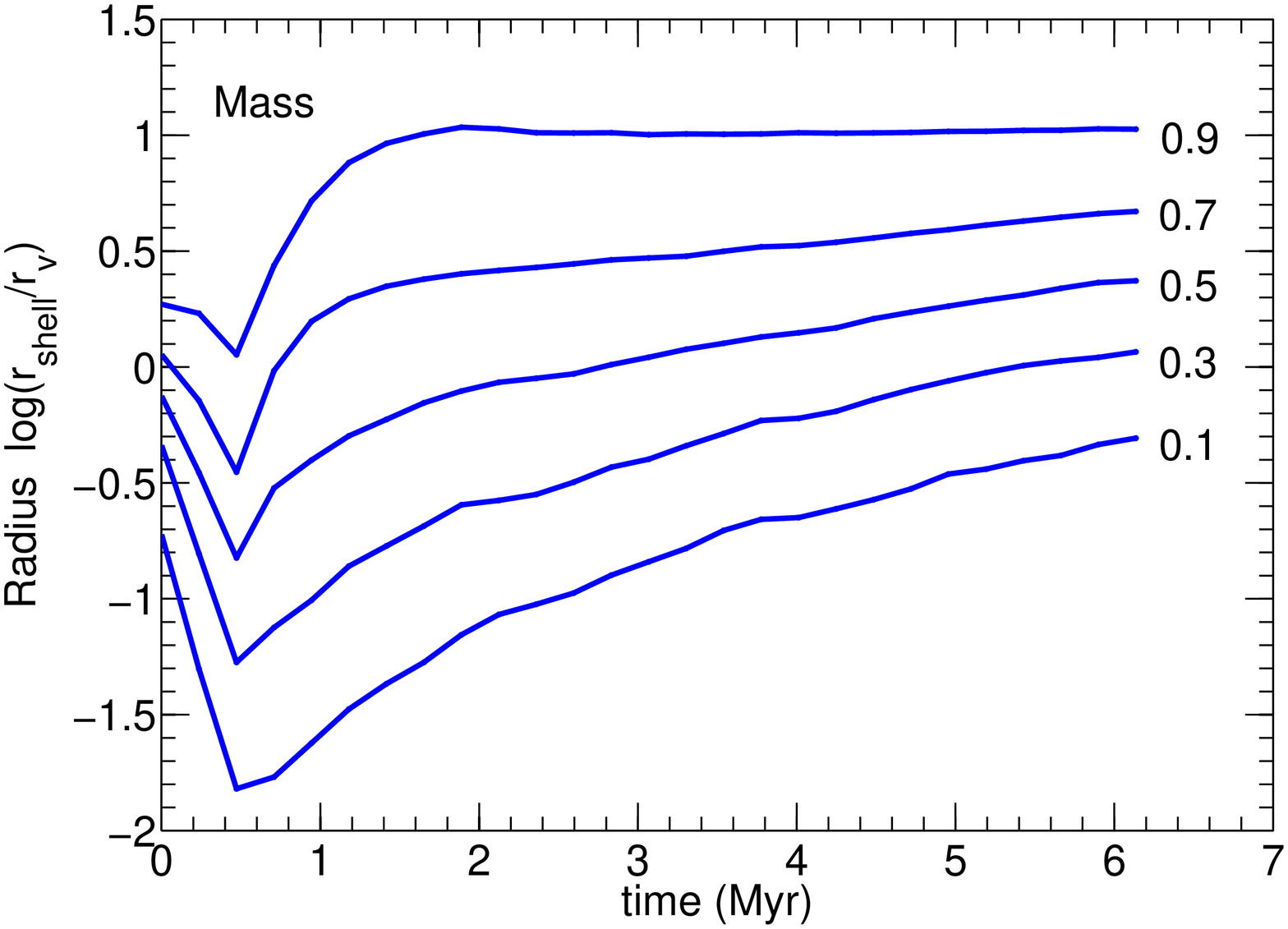}
    \includegraphics[width=6cm,angle=270]{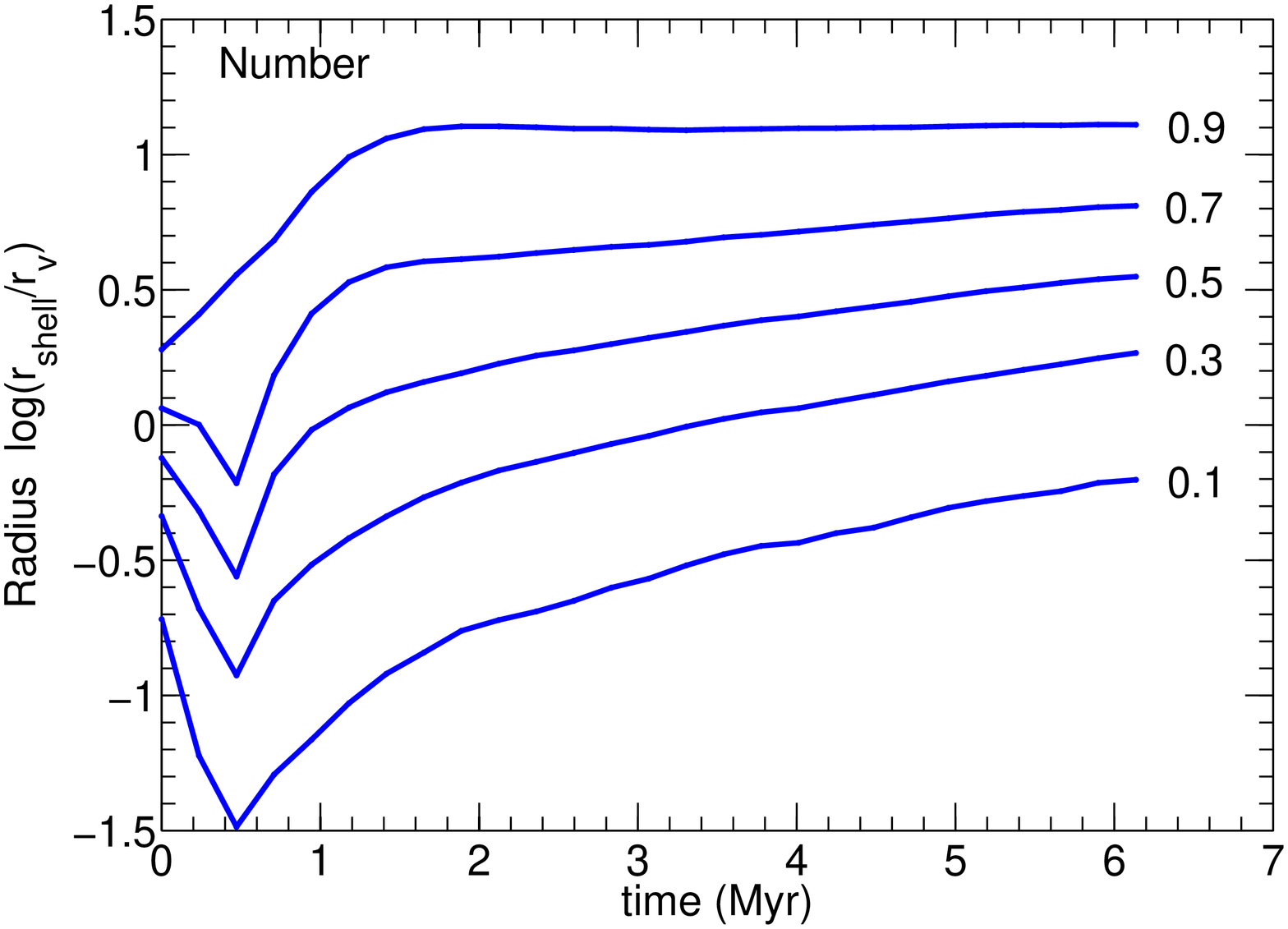}
  \end{center}
  \caption{Evolution of (left) mass and (right) number shells for
    clusters truncated at $5 r_{\rm v}$. The top and bottom
    panels are characterized by uniform and radius-dependent velocity
    dispersions, respectively. The numbers in the panels refer to the
    fractions (by number) of cluster stars included in the individual
    curves: the left-hand panels show increasing fractions of `most
    massive' stars from bottom to top, while the right-hand panels
    show the equivalent behavior for increasing fractions of `inner
    cluster' stars.
  \label{fig:shell_radius}}
\end{figure}

We define the mean separation between stars using the MST length,
\begin{equation}
\langle l \rangle=\frac{\rm MST}{n-1},
\end{equation}
where $n$ is the number of stars in the sample of interest. The mean
separation of a sample of randomly selected stars in the cluster can
be a trace of the mean separation within the
cluster. Fig. \ref{fig:MST-t} shows that the mean separation in the
cluster and of the 10 most massive stars varies with time.
%
%
It seems that for the radius-dependent velocity distribution, the
curve of the mean separation of massive stars forms a valley at
approximately 0.5 Myr, while the curve of the mean separation of the
full cluster increases from the onset. This valley corresponds to core
collapse of (only) the most massive stars in the cluster. A series of
radii of different mass shells (increasing fractions of `most massive'
stars, from bottom to top) is presented in the left-hand panels of
Fig. \ref{fig:shell_radius}. The right-hand panels of
Fig. \ref{fig:shell_radius} show increasing fractions of `inner
cluster' stars from bottom to top.
%
%
The evolution of the half-mass radius indicates that the cluster is
undergoing a core-collapse process (driven by the most massive stars)
until approximately 0.5 Myr, when it begins to re-expand. We compare
this timescale to the crossing time, $t_{\rm cross}$, which is simply
the cluster radius divided by its velocity dispersion.  For our
$N=1000, D=2.0$, and $q=0.30$ initial conditions, $t_{\rm cross} \sim
0.2$--0.6 Myr for the cluster as a whole. That is, core collapse of
the few most massive stars happens during the first few crossing
times.  However, the crossing time for the core region depends on the
adopted form for the radial velocity dispersion: it is 0.1--0.3 Myr
for the uniform distribution and 0.08--0.2 Myr for the
radius-dependent velocity dispersion.

For the radius-dependent velocity distribution, expansion of the outer
parts of the cluster after formation can be traced from the evolution
of the radius containing 90\% (by number) of cluster stars. Therefore,
the large $\Lambda'$ seen at early times in Fig. \ref{fig:Lambda-t} is
caused by expansion of the cluster rather than by a concentration of
the massive stars. This expansion is caused by the initially large
velocity in the central parts. Therefore, the radius-dependent
velocity dispersion would increase the velocity in the central part
without changing the initial virial ratio. On the other hand, the
evolution of the MST length of the cluster as a whole, which assigns
all members of a sample equal weights, can also be a tool for
measuring the size of the cluster.

\subsection{Mass segregation in velocity space}

The massive stars are becoming centrally concentrated very rapidly,
thus leading to rapid dynamical mass segregation. On the other hand,
stars of different masses may also exhibit different velocity
dispersions. We adopt our MST method here to measure mass segregation
in velocity-dispersion space. The only difference is that `distance'
now refers to the separation in velocity dispersion of two stars.
Velocity dispersion is defined as
\begin{equation}
\mbox{\boldmath $\sigma$}_i={\bf v_i}-{\bf v},
\end{equation}
where $\bm{\sigma}_i$ is the velocity dispersion of star $i$, ${\bf
v}_i$ the velocity of star $i$, and ${\bf v}$ the velocity of the
whole cluster. Therefore, the separation in velocity dispersion of two
stars can be expressed as
\begin{equation}
\mbox{\boldmath $\sigma$}_{ij}=\mbox{\boldmath
  $\sigma$}_i-\mbox{\boldmath $\sigma$}_j ={\bf v_i}-{\bf v_j},
\end{equation}
which is same as separation in terms of velocities.
%
\begin{figure}
  \begin{center}
    \includegraphics[width=9cm,angle=270]{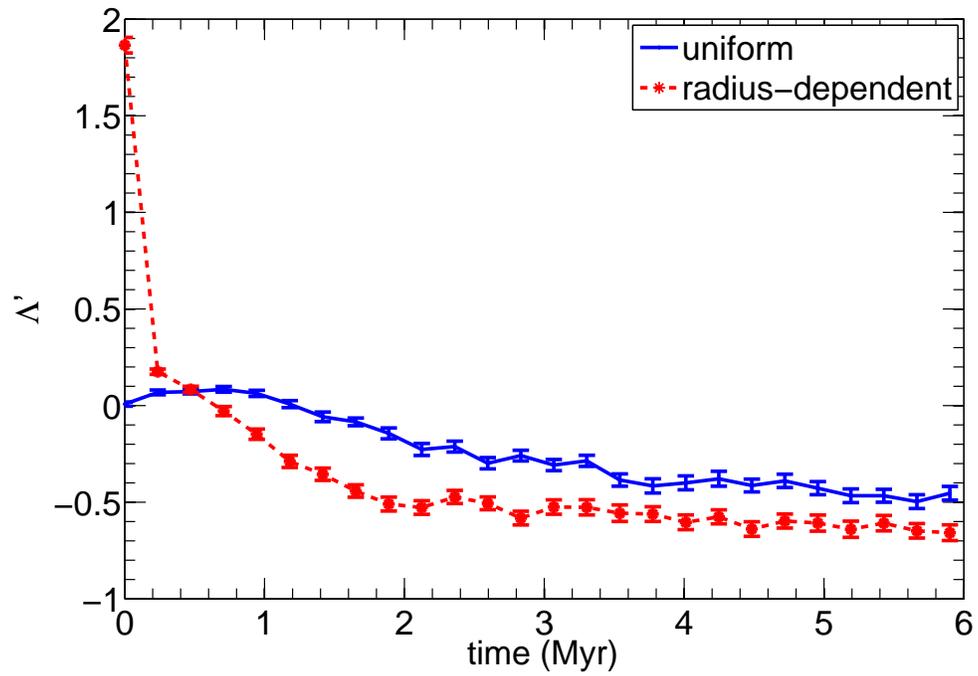}
  \end{center}
  \caption{Evolution of mass segregation in velocity space. The solid
    and dashed lines represent mass segregation for a uniform, random
    velocity dispersion and for the radius-dependent dispersion,
    respectively.
  \label{fig:MST_vspace}}
\end{figure}
%
From Fig. \ref{fig:MST_vspace} we find that clusters can exhibit
`inverse mass segregation' in velocity space by calculating $\Lambda'$
using the same treatment as before. The reason for this is that the
massive stars are preferentially concentrated in the core and have a
higher velocity dispersion than a sample of randomly selected stars.

\section{Conclusions}

We have presented simulations and analysis of the early evolution of
initially cool and clumpy star clusters by performing a large number
of $N$-body simulations. An MST method is used to measure the degree
of mass segregation. In addition, it can also be a ruler to trace
cluster size. Removal of the ejected stars is important physically,
since it has an effect on measuring the cluster's mass segregation.

We also conclude that the velocity distribution has an impact on mass
segregation during core collapse. For a radius-dependent velocity
distribution, which leads to a high velocity in the inner regions, the
picture of (the dynamical evolution of) the cluster during the
core-collapse period is very different from that resulting from the
assumption of a constant velocity dispersion. Although the high-mass
stars still undergo core collapse, the entire cluster expands from its
formation epoch. As a result, a high degree of mass segregation
results at the time of high-mass core collapse. We also adopt the MST
method to discuss mass segregation in velocity space, finding `inverse
mass segregation,' which may be caused by the
high velocity dispersion of the massive stars.

Because our aim in this paper was to explore the effects of a radially
dependent velocity dispersion on early mass segregation, we have not
included the effects of primordial binary stars (beyond their
dynamical formation in our simulations). This is clearly an important
omission, which we intend to address in our future work. A high {\it
initial} binary fraction would introduce, on average, a higher
velocity of escaping stars \citep{Weidner_Bonnel2010MNRAS}, which is
commonly seen in observations of young star clusters. However,
\cite{Weidner_Bonnel2010MNRAS} also found that the degree of mass loss
is nearly independent of whether or not the cluster possesses
primordial binaries. In addition, for a $N=1000$ cluster, varying the
binary fraction within reasonable bounds has little effect on the mean
mass of escapees. As a result, inclusion of initial binaries will
likely not significantly affect the signature of mass segregation in
our simulated clusters.

\section*{Acknowledgements}
We thank M. B. N. Kouwenhoven for valuable comments and
suggestions. JCY acknowledges partial financial support from the
National Natural Science Foundation of China (NSFC) through grants
11043006, 11043007 and 11073038. RdG acknowledges partial research support
through NSFC grants 11043006 and 11073001. LC is supported by NSFC
grants 11073038, Key Project 10833005, and by the 973
program: grant NKBRSFG 2007CB815403.

\bibliographystyle{apj}

\end{document}